\documentclass[twocolumn,showpacs,preprintnumbers,amsmath,amssymb]{revtex4}
\usepackage{graphicx}
\usepackage{amsmath}
\begin{document}

\title{Dynamics stabilization and transport coherency in a rocking ratchet for cold atoms}

\author{A.B. Kolton$^{1}$ and F. Renzoni$^{2}$}

\affiliation{$^{1}$CONICET, Centro Atomico Bariloche, 
8400 San Carlos de Bariloche, Rio Negro, Argentina}

\affiliation{$^{2}$Department of Physics and Astronomy, University College 
London, Gower Street, London WC1E 6BT, United Kingdom}

\date{\today}

\begin{abstract}
Cold atoms in optical lattices have emerged as an ideal system to 
investigate the ratchet effect, as demonstrated by several recent 
experiments. In this work we analyze theoretically two aspects of
ac driven transport in cold atoms ratchets. We first address the issue 
of whether, and to which extent, an ac driven ratchet for cold atoms can 
operate as a motor.
We thus study theoretically a dissipative motor for cold atoms, as obtained by 
adding a load to a 1D non-adiabatically driven rocking ratchet. We demonstrate 
that a current can be generated also in the presence of a load, e.g. the 
ratchet device can operate as a motor. Correspondingly, we determine the stall 
force for the motor, which characterizes the range of loads over which the 
device can operate as a motor, and the differential mobility, which 
characterizes the response to a change in the magnitude of the load. 
Second, we compare our results for the transport in an ac driven ratchet device 
with the transport in a dc driven system. We observe a peculiar phenomenon: 
the bi-harmonic ac force
stabilizes the dynamics, allowing the generation of uniform directed motion
over a range of momentum much larger than what is possible with a dc bias.
We explain such a stabilization of the dynamics by observing that
a non-adiabatic ac drive  broadens the effective cooling momentum range, and
forces the atom trajectories to cover such a region.  Thus the system
can dissipate energy and maintain a steady-state energy balance. 
Our results show that in the case of a finite-range velocity-dependent 
friction, a ratchet device may offer the possibility of controlling the 
particle motion over a broader range of momentum with respect to a purely 
biased system, although this is at the cost of a reduced coherency.
\end{abstract}

\maketitle

\section{Introduction}

The ratchet effect 
\cite{comptes,magnasco,adjari,bartussek,doering,reimann,rmp09} is usually 
defined as the rectification of 
fluctuations in the absence of a bias, as obtained in a system out of 
equilibrium. This typically corresponds to the generation of directed 
transport of particles through a macroscopically flat periodic structure.

The concept of ratchet leads naturally to the idea of Brownian motor, 
i.e. to the possibility of doing work against a load as a result of the 
rectification of fluctuations. Any ratchet device can then be characterized 
as a motor \cite{jaya04,linke,kostur,motor}, thus pointing out the features 
related to the possibility of doing work against a load. 

Cold atoms in optical lattices have emerged as an ideal system to investigate 
the ratchet effect \cite{advances,kastberg}. The significant advantage of this
system is tunability, which allows for arbitrary periodic potentials to be 
created, and drivings to be applied either under the form of rocking forces or
as modulation of the potential amplitude. This allowed for the realization of
1D rocking ratchets, both with periodic \cite{phil,gommers05a,gommers05b}
and quasiperiodic driving \cite{quasip}, and gating ratchets \cite{gating}. 
Furthermore, rectification mechanisms unique to higher-dimensional systems 
were demonstrated by using 2D rocking ratchets \cite{slav}.  
 Despite the number of successful realizations of driven ratchets for 
cold atoms, some fundamental aspects of the transport in these devices have 
yet to be analyzed. A first important issue is whether cold atom ratchets can,
and to which extent, operate as motors, e.g. can do positive work against a
load. A second fundamental issue is how the transport in cold atom ratchets 
compare with the transport as obtained by the direct application of a dc bias. 
%
This work precisely adresses these issues. First, we study and characterize 
theoretically the motor corresponding to a 1D rocking ratchet for cold atoms,  
as obtained by adding a load to the ratchet set-up. We demonstrate that a 
current can be generated also in the presence of a load, e.g. the ratchet 
device can operate as a motor. Correspondingly, we determine the stall force 
for the motor, which characterizes the range of loads over which the device 
can operate as a motor, and the differential mobility, which characterizes 
the response to a change in the magnitude of the load. Second, we compare 
transport in an ac driven ratchet device with the transport in a dc driven 
system. A quantitative study is performed, by determining the magnitude
of the obtained velocity, as well as the coherency of the directed transport.
It turns out that the ratchet set-up allows to generate uniform motion 
over a range of velocities larger than in the case of the dc driven system. 
An explanation for this {\it stabilization phenomenon} for ac driven 
transport is given. 

This work is organized as follows. In Sec. \ref{sec:setup} we describe 
the ratchet set-up. In Sec. \ref{sec:numerics} we first characterize a motor 
for cold atoms, as obtained by adding a load to an ac driven ratchet. We 
thus present our numerical results for the stall force and the differential 
mobility. We then determine the coherence of transport in an unbiased ratchet, 
and compare our results to a system of cold atoms in an optical lattice with a  
pure dc driving force. The observed stabilization phenomenon is discussed and 
analyzed.  In Sec. \ref{sec:conclusions} conclusions are drawn. 

\section{Ratchet set-up} \label{sec:setup}

The ratchet set-up considered in this work is the same as the one examined
in Ref.~\cite{brown}. A dissipative optical lattice is created by the 
interference of two counterpropagating laser fields, whose linear polarizations
are orthogonal (lin$\perp$lin  configuration \cite{robi}.) We consider 
a $J_g=1/2\to J_e=3/2$ atomic transition. This is the simplest atomic 
transition for which Sisyphus cooling \cite{cct} takes place. For this transition, the
light interference pattern results into a bipotential $U_{\pm}$
\begin{equation}
U_{\pm}(z) = \frac{U_0}{2}\left[ -2\pm \cos(2kz)\right]
\label{eq:pot}
\end{equation}
for the atoms, with the atom in the ground state sublevel $|+1/2\rangle$
experiencing the potential $U_{+}$, while the atoms in the other ground
state sublevel, $|-1/2\rangle$, will experience $U_{-}$. In Eq.~(\ref{eq:pot}), 
$U_0$ is the depth of the optical potential, $z$ the atomic position along the 
propagation direction of the laser fields, and $k$ the laser field 
wavevector. Besides the depth $U_0$ of the optical potential, another
important parameter to characterize the system is the photon scattering
rate $\Gamma^\prime{}$. The photon scattering rate from the excited state 
determines the stochastic optical transitions between the two ground state 
sublevels, with departure rates given by 
$\gamma_{\pm\to \mp}(z)=\Gamma^\prime{}(1\pm\cos(2kz))/9$. These transitions
lead to a friction force and to fluctuations in the atomic dynamics.

For the study of the ratchet effect, we include a bi-harmonic driving 
with frequencies $\omega_d$, $2 \omega_d$, and relative phase between
harmonics equal to $\phi$.  For an appropriate choice of the phase $\phi$ the
drive breaks the relevant time-symmetries of the system, and thus leads
to directed motion 
\cite{fabio,mahato,chialvo,dykman,goychuk,flach00,flach01,super,denisov08}.

In addition to the bi-harmonic ac drive,
we include a dc bias $F_{dc}$, which corresponds to the load for the motor.
The total applied driving force $F(t)$ is thus
\begin{equation}
F(t) = F_{ac} [ A_d \cos \omega_d t + B_d \cos(2 \omega_d t + \phi) ] + F_{dc}~.
\label{eq:driving}
\end{equation}
Here $F_{ac}$ is the amplitude of the ac drive, with relative amplitude 
between harmonics given by $A_d$, $B_d$, and $F_{dc}$ is the amplitude of the 
applied dc force.

As for the units, throughout this work, the atomic momenta will be expressed
in terms of the recoil momentum $p_r =\hbar k$ and the energies in terms of the
recoil energy $E_r=(\hbar k)^2/2m$, where $m$ is the atomic mass. The 
scattering rate and the time will be expressed in terms of the
recoil angular frequency $\omega_r=E_r/\hbar$, and its inverse.
Finally, the amplitudes of the applied forces will be expressed in terms of 
$F_r=\hbar k\omega_r$.

For all the results presented in this work, the following parameters are fixed: 
$U_0=200$E$_r$, and $\omega_d = \omega_v$, where $\omega_v$ is the
vibrational frequency of the atoms at the bottom of the wells. For the 
calculations with a bi-harmonic force we take: $A_d=B_d=1$, while 
$A_d=1$, $B_d=0$ holds when a single harmonic drive is considered for
comparison.

\section{Numerical results}\label{sec:numerics}

The numerical simulation techniques are the same as used in previous work \cite{brown}.
We simulate the atomic dynamics in the optical lattice in the presence
of the driving by semi-classical Monte Carlo simulations \cite{robi}.  
A second-order stochastic Runge-Kutta algorithm with adaptative time-step is 
used to evolve the atomic position and momentum between jumps corresponding
to changes of the internal state.
By averaging over the different atomic trajectories, it is then possible to 
determine the quantities of interest, as the average atomic momentum 
or the spatial diffusion coefficient, for different choices of parameters.
All numerical results presented in this work are producing by taking into
account $n = 10^4$ atomic trajectories. The atoms are initially at the bottom
of a well with zero momentum. The drive is then adiabatically turned on 
during 100 cycles. A further 1000 cycles are then used to calculate the quantities of 
interest.

The average steady-state atomic momentum is defined as
\begin{equation}
\langle p \rangle =\langle\lim_{t\to +\infty} \frac{1}{t} \int_0^t p(t)dt\rangle
\end{equation}
where $\langle \rangle$ indicates the average over the ensemble. In practice, 
in the simulation we calculate $ \langle p \rangle $ by averaging over the 
ensemble and over the last 1000 drive cycles, i.e. we calculate
$ \langle p \rangle  =  \overline{\langle p(t) \rangle} $, where the 
$\bar{~}$ sign indicates the time-average.

\subsection{Stall force and mobility}

\begin{figure}[hbtp]
\begin{center}
\includegraphics[height=2.in]{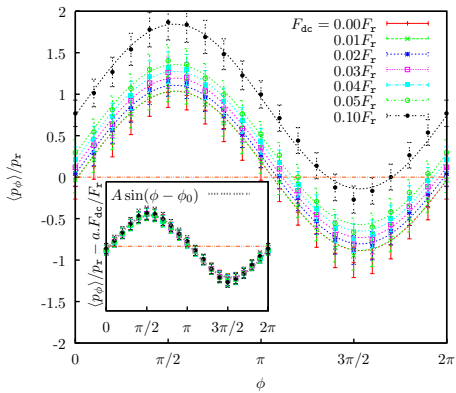}
\end{center}
\caption{(Color online) Average atomic momentum as a function of the relative 
phase $\phi$ between harmonics of the ac drive, for different values of the
applied dc bias. The data from our simulations are plotted together with the 
best fit with the function $\langle p_{\phi}\rangle/p_r = A\sin(\phi-\phi_0)$, 
where $A$ and $\phi_0$ are fitting parameters. The inset shows that by 
shifting the data by an amount $-\alpha F_{dc}/F_r$ proportional to the 
applied bias force, it is possible to overlap all the curves. Here $\alpha$ is 
an appropriately chosen constant. This proves that, for the examined range of
parameters, the dc bias results in a pure shift of the curve, whose
magnitude is proportional to the bias, with no significant distortion in shape.
The parameters of the calculation are: 
$\Gamma^\prime{}= 2\omega_r$, $F_{\rm ac} = 100 F_r$.
}
\label{fig:fig1}
\end{figure}

We first address the issue of whether, and to which extent, an ac driven 
ratchet for cold atoms can operate as a motor. 
We thus start our analysis by studying the atomic current as a function of the 
time-symmetry breaking parameter $\phi$ in the presence of a bias force 
$F_{dc}$. Our results are shown in Fig.~\ref{fig:fig1}.  

For zero bias the average atomic momentum has a sin-like dependence on the 
phase $\phi$, as  already extensively discussed in previous experimental 
and theoretical work on rocking ratchets for cold atoms 
\cite{phil,gommers05a,gommers05b,brown}, as well as in the framework of more 
general theory \cite{flach00,flach01,super}. For non-zero bias, the current 
still shows a sin-like dependence on the phase $\phi$ but with an additional 
vertical offset. Actually, as from the inset of Fig.~\ref{fig:fig1}, the 
overall shape of the current as a function of $\phi$ is essentially unchanged,
and the applied bias only produces an overall shift. Thus, the average momentum
is well described by the relationship:

\begin{equation}
\frac{\langle p_{\phi}\rangle}{p_r} = A \sin(\phi-\phi_0) + B~,
\label{eq:current1}
\end{equation}
where the phase $\phi_0$ corresponds to symmetry-breaking produced by
dissipation \cite{flach01,gommers05b}. Although $\phi_0$ is negligeable 
for the parameters of Fig. ~\ref{fig:fig1}, it can be large for other sets 
of parameters, as investigated in detail experimentally and theoretically 
in previous work \cite{gommers05b,brown,advances}.

Figure ~\ref{fig:fig1} proves that for any phase $\phi\neq \phi_0+n\pi$, with
$n$ integer, the ratchet can operate as a Brownian motor, i.e. can do work 
against a load, whose sign and maximum value depends on the phase $\phi$.
To characterize further this behaviour, we now define a specific Brownian
motor by fixing the Hamiltonian. Specifically, we consider the case of 
$\phi=3\pi/2$, so to break all the relevant symmetries. 

A first important parameter which characterizes a Brownian motor is the 
stall force $F_{stall}$. This is the largest bias force - or load - against 
which the motor can operate.

\begin{figure}[hbtp]
\begin{center}
\includegraphics[width=3.in]{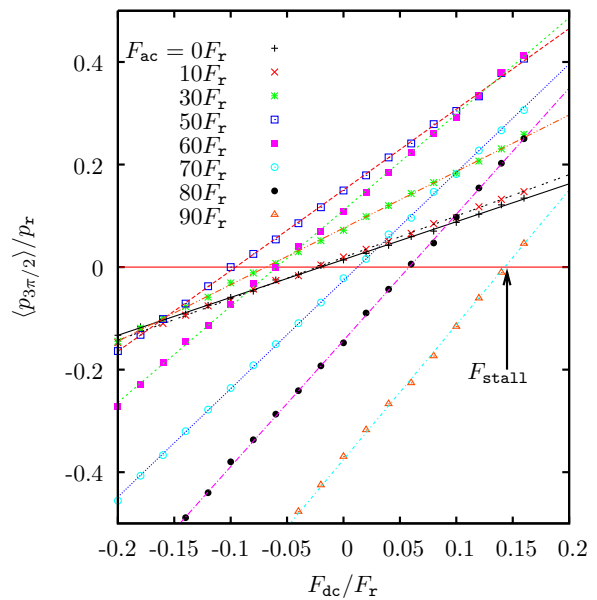}
\end{center}
\caption{(Color online) Load-velocity curve for the considered 1D rocking 
ratchet for cold atoms. The average atomic momentum is plotted as a function 
of the applied dc bias, for different amplitudes of the ac drive. The lines
are the best linear fits to the numerical data.}
\label{fig:fig2}
\end{figure}

To determine $F_{stall}$ we study the average atomic momentum 
$\langle p_{3\pi/2}\rangle$ as a function of the bias force amplitude $F_{dc}$.
We obtain in this way the so-called load-velocity curve, as reported in 
Fig.~\ref{fig:fig2} for different values of the driving strength. The crossing
point of the load-velocity curve with the horizontal axis defines the 
stall force $F_{stall}$. This is the maximum (in absolute value) load
against which the motor can do (positive) work.

\begin{figure}[hbtp]
\begin{center}
\includegraphics[width=3.in]{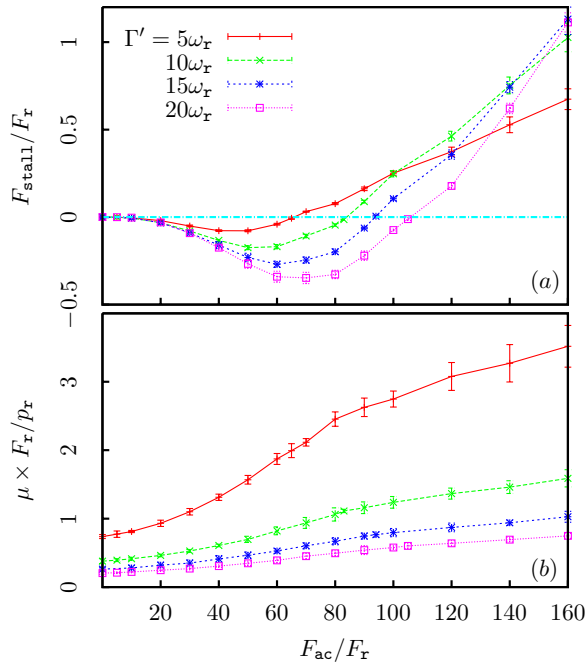}
\end{center}
\caption{(Color online) The stall velocity $F_{stall}$ (a) and the mobility
$\mu$ (b), as derived from plots as those in Fig. \protect\ref{fig:fig2}, are
plotted as a function of the amplitude of the ac drive, for different
values of the scattering rate $\Gamma^\prime{}$.
}
\label{fig:fig3}
\end{figure}

The results from Fig.~\ref{fig:fig2} show a linear relationship between the
average atomic momentum and the applied dc bias, i.e. the (differential) 
mobility $\mu$
\begin{equation}
\mu = \frac{d\langle v(F_{dc})\rangle}{dF_{dc}}
\label{mobility}
\end{equation}
is independent of $F_{dc}$. Thus, the relationship between the
average atomic momentum and the applied dc bias can be well described by:

\begin{equation}
\frac{\langle p_{3\pi/2} \rangle}{p_r} = m\omega_r \mu(F_{ac},\Gamma^\prime{})
\left( F_{dc}-F_{stall}(F_{ac},\Gamma^\prime{})\right)/F_r
\label{eq:current2}
\end{equation}

where we evidenced that both $F_{stall}$ and the mobility $\mu$ depend
on the strength of the ac drive as well as on the scattering rate
$\Gamma^\prime{}$. The dependence of these parameters on $F_{ac}$ is shown in
Fig.~\ref{fig:fig3}. The two parameters show a very different dependence
on the strength of the driving. 

The stall force, which characterizes the
range over which the ratchet device can operate as a motor, is a
nonmonotonic function of $F_{ac}$ (see Fig.~\ref{fig:fig3}), and changes sign
for increasing $F_{ac}$. This behaviour is similar to the nonmonotonic 
dependence of the atomic current on the strength of the ac drive in the 
absence of a dc bias, as studied in Refs.~\cite{phil,brown} (see also 
Fig.~\ref{fig:fig4}(a)).   In particular, the changes of sign
of $F_{stall}$ correspond to current reversals in the unbiased
ratchet. The similarity between the two curves can be simply explained 
in terms of Eq.~\ref{eq:current2}, and noticing that for our system 
the mobility is a positive monotonic function of $F_{ac}$, as discussed now.

The atomic mobility $\mu$ characterizes the atomic velocity in response to a 
change in dc bias or load as from the definition of Eq.~\ref{mobility}. Our results of
Fig.~\ref{fig:fig3} show that the mobility is a positive increasing function
of the strength of the ac drive, i.e. the same change in dc force results in
a larger average increment in atomic momentum for a larger ac drive. This 
behaviour is enhanced at 
small scattering rate $\Gamma^\prime{}$. As anticipated above, as $\mu$ is a 
monotonic positive function of $F_{\rm ac}$, the stall force $F_{\rm stall}$
and $\langle p_{3\pi/2}\rangle$, related via the mobility - see 
Eq.~\ref{eq:current2}, shows a similar dependence on $F_{\rm ac}$.

We also notice that combining Eqs.~(\ref{eq:current1}, \ref{eq:current2}) we 
obtain an expression for the average atomic momentum as a function of $\phi$:

\begin{equation}
\frac{\langle p_{\phi} \rangle}{p_r} = 
\frac{m\omega_r \mu}{\cos\phi_0} F_{stall}\sin(\phi-\phi_0)
+ m\omega_r \mu F_{dc}
\end{equation}

which depends only on the mobility $\mu$, the stall force $F_{stall}$ and the
phase-shift $\phi_0$ determined by dissipation. It is thus possible to express 
the atomic current in the general case, i.e. with or without a load, as a 
function of the parameters $\mu$, $F_{stall}$ which fully characterize the 
operation of the atomic device with a load. 

\subsection{Coherency of the atomic transport}

We now turn to the quantitative characterization of transport in an 
(unbiased) ac driven ratchet. This will then be used to compare ac driven 
transport to transport induced by an applied dc bias. 

The coherency of the directed transport in a ratchet device can be
characterized by comparing the linear transport to the spatial spread. 
Quantitatively, a standard measure \cite{linke,machura,roy} of the coherency 
is the Peclet number
\begin{equation}
{\rm Pe} = \frac{|\langle v\rangle |l}{D_{\rm sp}}~.
\end{equation}
Here $l$ is a characteristic length of the system, which in our case will be
taken equal to $l=\lambda = 2\pi/k$, twice the spatial period of the optical 
lattice. $D_{sp}$ is the spatial diffusion coefficient, defined as
\begin{equation}
D_{\rm sp} = \lim_{t\to +\infty} \frac{\langle x^2(t) \rangle - 
\langle x(t) \rangle^2}{2 t}~. 
\end{equation}

\begin{figure}[hbtp]
\begin{center}
\includegraphics[width=3.in]{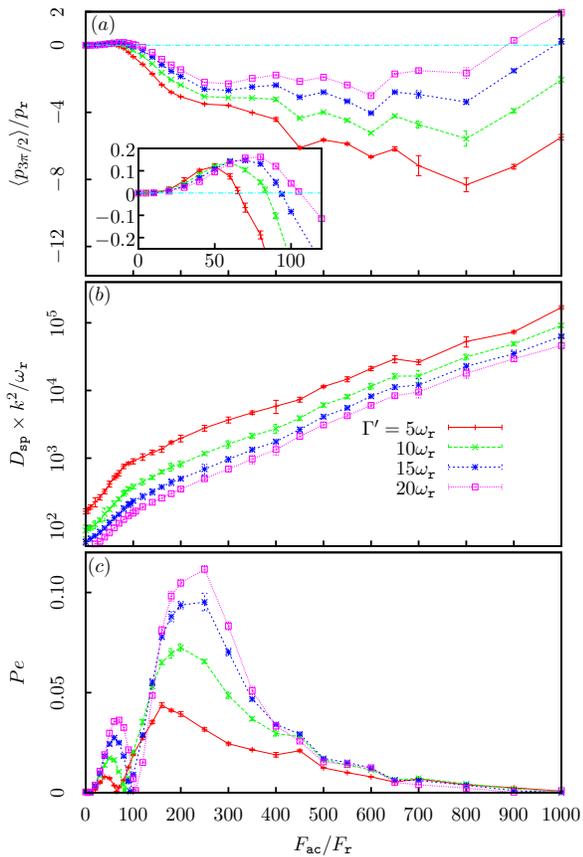}
\end{center}
\caption{(Color online) Study of the coherency of the cold atom ratchet. 
The average atomic momentum (a), the spatial diffusion coefficient (b) and 
the resulting Peclet number are reported as a function of the amplitude of 
the ac drive for different values of the scattering rate $\Gamma^\prime{}$.
All the data refers to an unbiased system, i.e. $F_{dc}=0$.}
\label{fig:fig4}
\end{figure}

Our results for $\langle p\rangle = m \langle v\rangle$, the spatial diffusion
coefficient $D_{\rm sp}$ and the resulting Peclet number are shown in 
Fig.~\ref{fig:fig4}.  In (a) the average momentum is plotted as a function of
the ac drive amplitude, at different values of the bias force. The regime of 
small driving, magnified in the inset, was already examined in 
Refs.~\cite{phil,brown}. There, the nonmonotonic dependence of the average 
momentum on the strength of the ac drive was discussed in details, and it 
was observed that the current first increases (in absolute value) 
with $F_{ac}$, then decreases and changes sign. Out present study extends
well beyond the regime analyzed previously, and shows that beyond
the previously observed reversal, the current increase significantly. It 
then reaches a maximum and start decreasing. The data shows evidence of a 
second current reversal.

In Fig.~\ref{fig:fig4}(b) we report our numerical results for the spatial 
diffusion coefficients for a driven optical lattice \cite{regis}. Our data
show that the spatial diffusion coefficient is a rapidly increasing function of the
strength of the ac drive, and this behaviour is enhanced at small values of
the scattering rate $\Gamma^\prime{}$. By combining the data for the atomic 
momentum (Fig.~\ref{fig:fig4}(a)) and for the spatial diffusion coefficient
(Fig.~\ref{fig:fig4}(b)) we derived the Peclet number which characterizes the
coherency of the transport, as plotted in Fig.~\ref{fig:fig4}(c). The Peclet
number shows two (local) maxima as a function of $F_{\rm ac}$. The first 
maximum is observed in the regime of small driving forces. In this regime
also the average momentum shows a maximum, i.e. the maximum in the coherency 
is determined both by the nonmonotonic dependence of the average momentum
and by the monotonic increase of the diffusion coefficient. The second maximum
of the coherency is observed in the range of forces beyond the first current 
reversal. In this case around the maximum of the coherence the average 
momentum is an increasing function of $F_{\rm ac}$, and the maximum in 
coherency is due to the faster monotonic increase in the diffusion coefficient.
The largest value for the coherency is associated with the second maximum. We 
notice, however, that although the average momentum for the second maximum is 
about one order of magnitude larger than the momentum for the first maximum,
the increase in coherency is only of the order of a factor three, because of 
the large increase in the diffusion coefficient.


\subsection{Dynamics stabilization}

\begin{figure}[hbtp]
\begin{center}
\includegraphics[width=3.in]{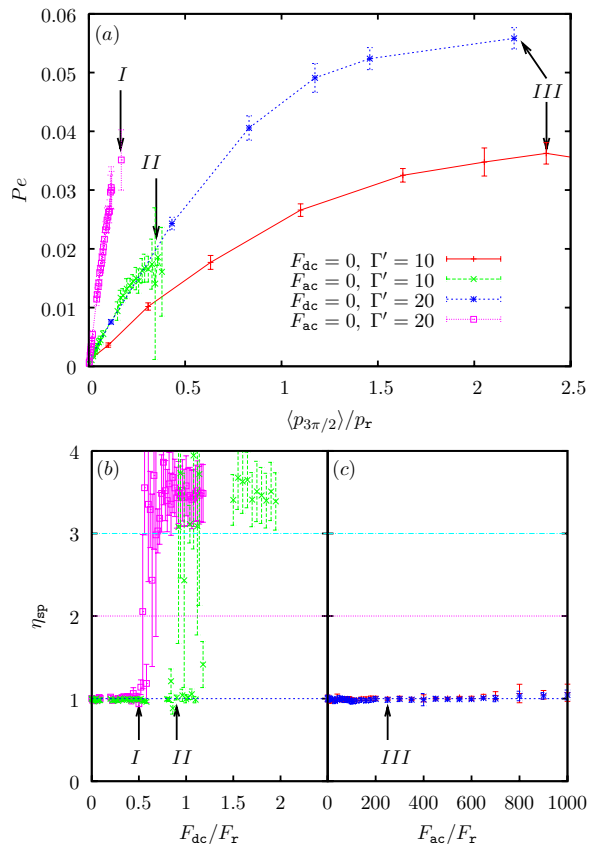}
\end{center}
\caption{(Color online) Comparison between the coherency of transport
in a cold atom ratchet and the transport produced by applying a dc force.
In (a) the Peclet numbers for the trasport produced by an applied dc force to
cold atoms in an optical lattice are reported together with the Peclet
number for the trasport produced by a applied ac bi-harmonic drive. In both
cases the Peclet number is plotted as a function of the obtained average
atomic momentum. The data for the ac-drive case were generated by considering
$F_{ac}$ values from the first current reversal to the maximum of Pe (see
Fig.~\protect\ref{fig:fig4}). In (b) the spatial diffusion esponent is
plotted as a function of the dc force, with $F_{ac} = 0$. In (c) the spatial
diffusion esponent is plotted as a function of the ac force amplitude, with
$F_{dc} = 0$. To allow for comparison between (b) and (c) with (a), for each
of the curves in (b) and (c) we indicate with an arrow the driving strength 
(dc and ac, respectively) corresponding to certain values of the atomic 
momentum which are also reported in (a).}
\label{fig:fig5}
\end{figure}

As final point of our analysis, we aim to compare the characteristics of the
transport in the considered 1D rocking ratchet with the features for the 
transport obtained by simply applying a dc force. We therefore made 
additional numerical
simulations for atoms in an optical lattice in the presence of an applied 
dc force of amplitude $F_{dc}$, without ac driving ($F_{ac}=0$). The results
of our numerical calculations are shown in Fig.~\ref{fig:fig5}, 
for two different scattering rates $\Gamma^\prime{}$.
In order to be able to compare the coherency for the two set-ups, we plotted
the coherency as a function of the generated momentum. In this way, for a
given generated momentum we can meaningfully compare the coherences for 
transport induced by a dc bias and by ac forces.

Fig.~\ref{fig:fig5}(a) shows that 
for both the considered values of $\Gamma^\prime{}$, the
transport induced by the dc bias shows a larger coherency than the one 
generated by the ratchet effect. This is due to the fact that, as we 
verified numerically, for a dissipative optical lattice of the type considered 
here an applied dc bias does not modify significantly the diffusion 
coefficient \cite{note}, at variance with the large increase produced by an 
ac driving.  Thus, a given average atomic momentum can be induced with a larger
coherency by a dc force rather than by an ac driving. However, as an important
point, we notice that by applying a dc force it is possible to generate uniform
motion only over a limited range of atomic momentum, and this is the reason why 
the curves in Fig.~\ref{fig:fig5} corresponding to the dc-bias case extend 
only up to a certain value of the atomic momentum. Beyond that value, the 
motion becomes accelerated. To verify this point, we determined from our data
the diffusion exponent $\eta_{sp}$ defined as
\begin{equation}
\langle x^2(t)\rangle - \langle x(t)\rangle^2 \sim t^{\eta_{sp}}
\end{equation}
in the limit $t\to \infty$. According to the definition,  $\eta_{sp}=1$
corresponds to normal diffusion, and $\eta_{sp}=4$ to uniformly accelerated 
motion. It is therefore a sensitive quantity to detect persistent accelerated motion.
Our results for the diffusion exponent are reported in Fig.~\ref{fig:fig5}(b) 
and (c) for the case of dc and ac driving, respectively. Fig.~\ref{fig:fig5}(b)
shows clearly that the application of a dc bias beyond a certain crytical value
results into a crossover to accelerated motion, characterized by $\eta_{sp}=4$.
The actual numerical value is somewhat smaller than $\eta_{sp}=4$ for a finite
time window indicating that  not all the atoms may be accelerated just above 
the threshold. In order to understand this behaviour, it is sufficient to 
recall that the damping mechanism associated to Sisyphus cooling has a finite 
capture range \cite{cct}. Thus, above a certain threshold an applied dc force 
will accelerate most of the atoms. This behaviour has to 
be contrasted to the dynamics produced by ac forces. Figure~\ref{fig:fig5}(a) 
shows that the use of an ac drive allows the generation of uniform motion
over a large range of momentum, much larger than what possible with a dc bias.
This is confirmed by Fig.~\ref{fig:fig5}(c) which shows that the diffusion is 
normal over a very broad range of applied ac forces. Thus, we observe a 
surprising phenomenon: the bi-harmonic ac force stabilize the dynamics, 
allowing the generation of uniform directed motion over a range of momentum
much larger than what possible with a dc bias.

\begin{figure}[hbtp]
\begin{center}
\includegraphics[width=3.in]{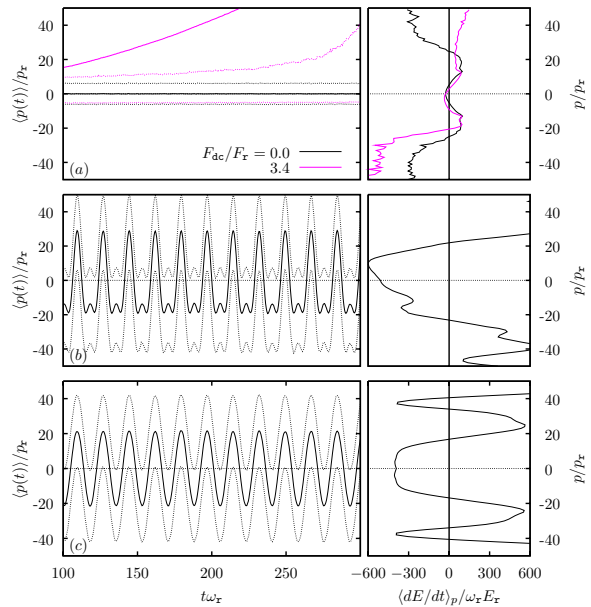}
\end{center}
\caption{(Color online) Ensemble-averaged time-dependent atomic momentum 
(solid line) as a function of time. 
Left: In (a) the case of a dc drive with $F_{dc}=3.4 F_r$ is considered together
with the unbiased case $F_{dc}=0$; 
(b) refers to a bi-harmonic drive with $F_{ac}= 600F_r$. Finally,
(c) report the data for a single-harmonic drive with $F_{ac}=600F_r$. 
In all cases the dashed lines indicate the instanteneous interval 
$[p_{-}:p_{+}]$  which contains 68\% of the momentum distribution,
with $P(p>p_{+}) = P(p<p_{-})=16$\%. We notice that for (a) $p_{-}$ 
is, within the numerical accuracy, the same for the two considered 
cases.
Right: The plots show the time-average derivative of the total 
atomic energy as a function of the atomic momentum.
}
\label{fig:fig6}
\end{figure}

To explain such a stabilization of the dynamics, we refer
to Fig.~\ref{fig:fig6}, where the ensemble-averaged time-dependent 
atomic momentum is studied for three different situations: dc drive, 
bi-harmonic driving, and single harmonic drive. The dc drive case also
includes a calculation for an unbiased system for comparison. The 
time-derivative of the total atomic energy 
\begin{equation}
E =\frac{p^2}{2m}+U_{\pm}(x)
\end{equation}
is also reported for the three cases as a function of the atomic momentum, 
to identify the ranges of atomic momentum in which dissipation takes place.
The derivative $\langle dE(p)/dt\rangle $ is calculated from the atomic 
trajectories in the following way. For a given mometum $p$ we identify 
in the considered $n$ atomic trajectories the $N(p)$ instants $t_i$
($i=1...N$) at which an atom has a momentum $p$. We then determine the 
energy increment $\Delta E_i$ following the (variable step-size)
increment $dt_i$. The time-derivative is then given by 
\begin{equation}
\langle dE(p)/dt\rangle = \frac{1}{N(p)}\sum_{i=1}^{N(p)} 
\frac{\Delta E_i}{dt_i}~.
\end{equation}
This quantity is the average power input of the external forces, noise
and internal state transitions which are responsible for the cooling and
heating processes.
Consider first the case of an unbiased and undriven system, as in
Fig.~\ref{fig:fig6}(a). We can distinguish a range of momentum in 
which energy is in average dissipated ({\it cooling region}):
$\langle dE(p)/dt\rangle <0 $, and a
{\it heating region} in which energy increases. At steady state,
the atomic momentum distribution extends over both regions, so that the
average energy is constant. We notice that atoms explore the heating 
region to then return to the cooling region, i.e. there is a continuous 
interchange of atoms between the two regions. Consider now the application 
of a dc force. The corresponding time-derivative of the energy is then 
modified by a term which includes the power supplied by the bias. As from 
Fig.~\ref{fig:fig6}(a) beyond a certain threshold
(as identified in Fig.~\ref{fig:fig5}), the atoms are accelerated. This 
is because essentially a large fraction of the momentum distribution is now 
located in the heating region, and the atoms can persist there. As a given atom
trajectory does not cover the cooling region,  energy is not dissipated and 
the steady-state energy balance is lost. Accelerated motion is thus produced.

Consider now the application of ac drivings. Both cases of bi-harmonic 
and single-harmonic drive are studied in Fig.~\ref{fig:fig6} (b) and (c),
respectively. As already obseved,   bi-harmonic drive allows the 
generation of uniform motion over a range of momentum much larger than 
in the case of a dc bias. Such a surprising stabilization of the dynamics 
can be explained as a result of two contributions. First, an ac drive broadens 
the effective cooling region. Second, for an ac drive the time-dependent 
atomic momentum oscillates in time. This forces the atom trajectories to cover
the cooling region.  Thus the system can dissipate energy and maintain an
energy balance. As from Fig.~\ref{fig:fig6} (b) and (c), this holds for both
bi-harmonic and single-harmonic drivings. Furthermore, for a bi-harmonic drive
the energy balance allows for the generation of directed motion at constant 
average velocity over a broad range of atomic momentum.

\section{Conclusions}\label{sec:conclusions}

In conclusion, in this work  we analyzed two aspects of
ac driven transport in cold atoms ratchets. We first addressed the issue
of whether, and to which extent, an ac driven ratchet for cold atoms can
operate as a motor. We thus studied  and characterize theoretically 
the motor corresponding to a 1D rocking ratchet for cold atoms,  as obtained 
by adding a load to the ratchet set-up. We demonstrated that a current can be 
generated also in the presence of a load, e.g. the ratchet device can operate
as a motor. Correspondingly, we determined the stall force for the motor, 
which characterizes the range of loads over which the device can operate as a
motor, and the mobility, which describes the sensitivity to changes in the 
load magnitude.  We then compared our results
for the transport in an ac driven ratchet device with the transport in a dc
driven system. We observed a peculiar phenomenon: the bi-harmonic ac force 
stabilizes the dynamics, allowing the generation of uniform directed motion 
over a range of momentum much larger than what possible with a dc bias.
We explained such a stabilization of the dynamics by observing that 
a non-adiabatic ac drive  broadens the effective cooling momentum
range, and 
forces the atom trajectories to cover such a region.  Thus the system 
can dissipate energy and maintain an energy balance.  This allows for 
the generation of directed motion at constant average velocity over a 
broad range of atomic momentum.

As an interesting point, our results show that in the case of a finite-range 
velocity-dependent friction, a ratchet device may offer the possibility 
of controlling the motion over a broader range of momentum with respect to 
a purely biased system, although this is at the cost of a reduced coherency.

\section{Acknowledgements}

We thank the Royal Society and the Leverhulme Trust for financial support.

\end{document}